\documentclass{article}
\usepackage{spconfa4,amsmath,graphicx}
\usepackage{amsmath,amsfonts}
\usepackage{algorithmic}
\usepackage{algorithm}
\usepackage{array}
\usepackage{hyperref}
\usepackage[caption=false,font=normalsize,labelfont=sf,textfont=sf]{subfig}
\usepackage{textcomp}
\usepackage{stfloats}
\usepackage{url}
\usepackage{verbatim}
\usepackage{graphicx}
\usepackage{cite}
\usepackage{bm}
\usepackage{multirow}
\usepackage{amssymb}
\usepackage{comment}
\usepackage{mathtools}
\usepackage{booktabs}
\usepackage{balance}
\usepackage{flushend}
\usepackage{siunitx}
\usepackage{xcolor}
\usepackage{arydshln}
\AtBeginDocument{
  \abovedisplayskip     =1\abovedisplayskip
  \abovedisplayshortskip=1\abovedisplayshortskip
  \belowdisplayskip     =1\belowdisplayskip
  \belowdisplayshortskip=1\belowdisplayshortskip
}

\def\R{{\mathbb R}}

\def\thineq{\hspace{-.1em}=\hspace{-.1em}}

\makeatletter
\def\bstctlcite{\@ifnextchar[{\@bstctlcite}{\@bstctlcite[@auxout]}}
\def\@bstctlcite[#1]#2{\@bsphack
  \@for\@citeb:=#2\do{%
    \edef\@citeb{\expandafter\@firstofone\@citeb}%
    \if@filesw\immediate\write\csname #1\endcsname{\string\citation{\@citeb}}\fi}%
  \@esphack}
\makeatother 

\let\OLDthebibliography\thebibliography
\renewcommand\thebibliography[1]{
  \OLDthebibliography{#1}
  \setlength{\parskip}{.5pt}
  \setlength{\itemsep}{1pt plus 0.3ex}
}

\title{TF-LOCOFORMER: TRANSFORMER WITH LOCAL MODELING BY CONVOLUTION\\FOR SPEECH SEPARATION AND ENHANCEMENT}
\name{Kohei Saijo$^{1,2*}$, Gordon Wichern$^{1}$, François G. Germain$^{1}$, Zexu Pan$^{1}$, Jonathan Le Roux$^{1}$ \thanks{*This work was done during an internship at MERL.}}
\address{$^1$Mitsubishi Electric Research Laboratories (MERL), Cambridge, MA, USA \;  $^2$Waseda University, Japan}
\begin{document}
\bstctlcite{IEEEexample:BSTcontrol} %

\ninept
\maketitle
\begin{abstract}
Time-frequency (TF) domain dual-path models achieve high-fidelity speech separation. While some previous state-of-the-art (SoTA) models rely on RNNs, this reliance means they lack the parallelizability, scalability, and versatility of Transformer blocks. Given the wide-ranging success of pure Transformer-based architectures in other fields, in this work we focus on removing the RNN from TF-domain dual-path models, while maintaining SoTA performance. This work presents TF-Locoformer, a Transformer-based model with LOcal-modeling by COnvolution. The model uses feed-forward networks (FFNs) with convolution layers, instead of linear layers, to capture local information, letting the self-attention focus on capturing global patterns. We place two such FFNs before and after self-attention to enhance the local-modeling capability. We also introduce a novel normalization for TF-domain dual-path models. Experiments on separation and enhancement datasets show that the proposed model meets or exceeds SoTA in multiple benchmarks with an RNN-free architecture.
\end{abstract}
\section{Introduction}
\label{sec:introduction}

The past decade has witnessed dramatic progress in speech separation thanks to advancements in neural networks (NNs).
Early work such as deep clustering~\cite{dc} trains NNs to estimate time-frequency (TF) masks in the TF-magnitude domain.
In contrast, time-domain audio separation network (TasNet)-style NNs have improved separation performance drastically by introducing learnable encoders and decoders~\cite{convtasnet}.
Dual-path modeling, which chunks the features and conducts local and global modeling alternately for efficient sequence modeling, is now one of the mainstream approaches in time-domain end-to-end (E2E) networks~\cite{dprnn, dptnet, sepformer}.
More recently, dual-path modeling in the TF domain, where temporal and frequency modeling are done alternately, has shown impressive performance improvements~\cite{tfpsnet,tfgridnet}.
TF-domain models have the potential to perform better in realistic reverberant conditions, as the FFT window size is usually much longer than the kernel size of a typical learnable encoder~\cite{cord2022monaural}.

While Transformer-based architectures~\cite{transformer} have shown great success in time-domain E2E separation~\cite{sepformer, qdpn, mossformer2}, the current state-of-the-art (SoTA) TF-domain dual-path separation model relies on RNNs~\cite{tfgridnet}.
Although RNNs have some advantages such as smaller memory cost in inference,  the training is typically very time-consuming since it cannot be parallelized.
In contrast, Transformer-based models have been eagerly investigated in other fields~\cite{transformer, karita2019comparative, vit} because of their multiple advantages: the training process can be parallelized, models can accept prompts (e.g., to specify the task~\cite{whisper, uses}), and they may scale well~\cite{scaling_law, scaling_law_acoustic_model, scaling_law_vit}; these features are either unattainable or at least unconfirmed in RNNs.
Before realizing these potential benefits of Transformer-based models, the first step, which is the goal of this paper, is to investigate whether
we can obtain comparable or better performance as RNN-based SoTA models with an RNN-free model of similar complexity.

Global and local modeling often both play an important role in speech processing~\cite{conformer}. 
There are strong hints in the literature that this is also true for speech separation, and that Transformer blocks lack an intrinsic local modeling ability.
Indeed, Transformer blocks, which excel at global modeling thanks to self-attention, have worked well in the context of time-domain dual-path models~\cite{sepformer}, where self-attention in the intra-chunk path is limited to local modeling by construction, while failing to compete in TF-domain dual-path architectures, where local modeling is not explicitly enforced. %
On the other hand, RNNs, which can capture local information, do attain strong performance in TF-domain dual-path models. Notably, the current SoTA model in TF-domain separation, TF-GridNet~\cite{tfgridnet}, exploits both RNNs and self-attention. %
As an alternative to RNNs for capturing local information, we consider inserting convolution layers within a Transformer-based dual-path architecture.

\begin{figure}[t]
\centering
\centerline{\includegraphics[width=0.9\linewidth]{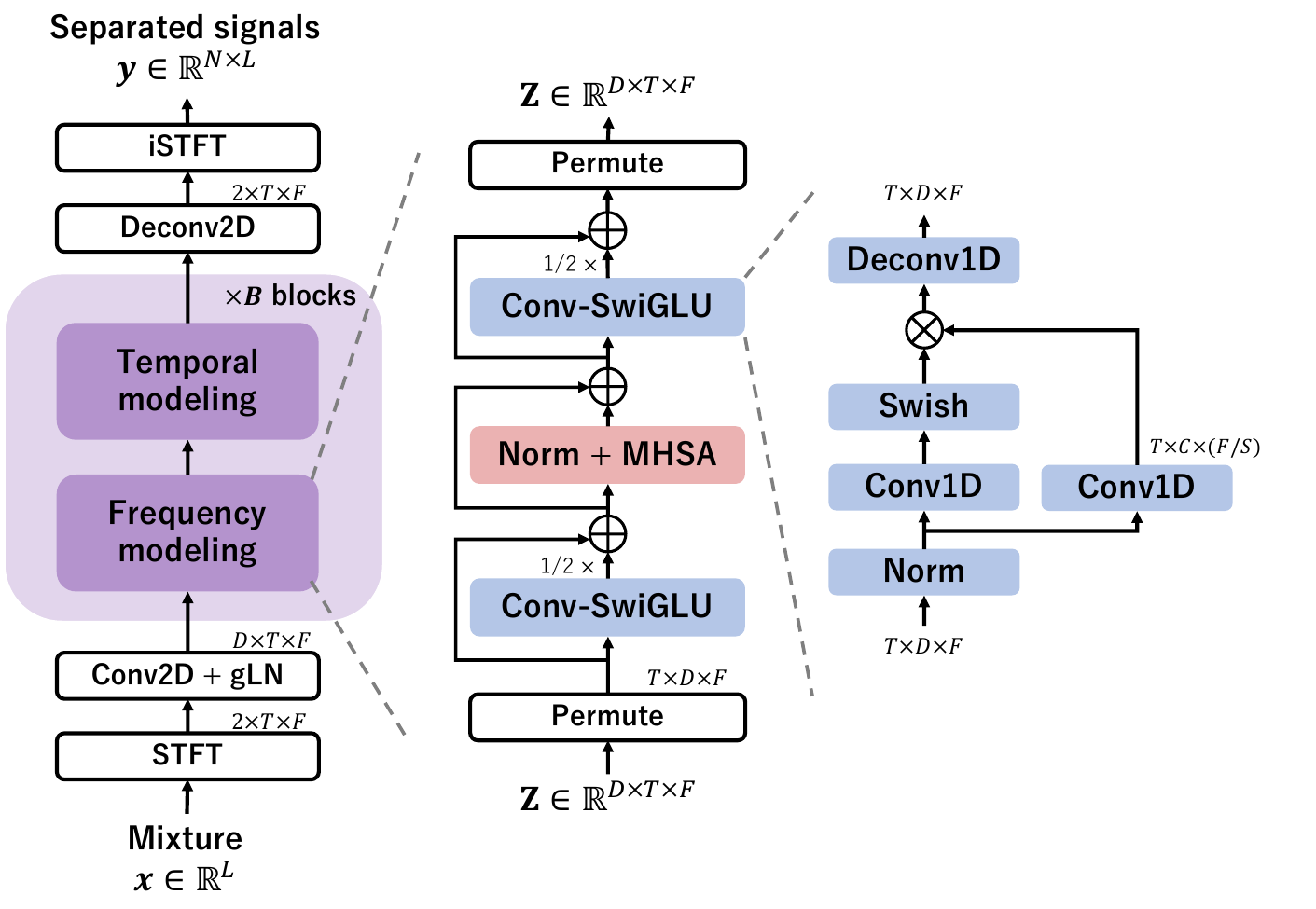}}\vspace{-2mm}
\caption{
    Overview of the proposed TF-Locoformer.
    The temporal modeling block is the same as the frequency modeling block with a permutation of the time and frequency dimensions.
}
\label{fig:overview}
\vspace{-4mm}
\end{figure}

We propose TF-Locoformer (TF-domain Transformer with LOcal modeling by COnvolution), a simple extension of the Transformer-based model that alternates global and local modeling.
Starting from the normal Transformer block, we first replace the two linear layers in the feed-forward network (FFN) with 1d-convolution and 1d-deconvolution layers, respectively. %
In addition, we leverage swish gated linear unit (SwiGLU) activations in the FFNs 
and place such FFNs both before and after self-attention, inspired by the success of the \textit{macaron-style} architecture~\cite{macaron_net, conformer}.
To further improve the performance, we introduce a novel normalization layer for TF-domain dual-path models.
We empirically demonstrate that these extensions boost the local-modeling capability and enable TF-Locoformer to achieve comparable or better performance than the current state-of-the-art RNN-based models.
Our source code is available online\footnote{\url{https://github.com/merlresearch/tf-locoformer}}.

\section{TF-Locoformer}
\label{sec:model}

\subsection{Overview of TF-Locoformer}
\label{ssec:overview}

Figure~\ref{fig:overview} shows the overview of the proposed TF-Locoformer model.
The model is based on TF-domain dual-path modeling~\cite{tfpsnet, tfgridnet}, where frequency and temporal modeling are done alternately. It separates each source by complex spectral mapping, where their real and imaginary (RI) components are directly estimated.

Let us denote a monaural mixture of $N$ speech signals $\bm{s}\in \R^{N \times L}$ and a noise signal $\bm{b} \in \R^{L}$ as $\bm{x} = \sum\nolimits_{n} \bm{s}_{n} + \bm{b} \in \R^{L}$, where $L$ is the number of samples in the time domain and $n=1,\dots,N$ is the speech source index.
In short-time Fourier transform (STFT) domain, the mixture is written as $\bm{X} \in \R^{2 \times T \times F}$, where $T$ and $F$ are the number of frames and frequency bins, and 2 corresponds to real and imaginary parts.

The input $\bm{X}$ is first encoded into an initial feature $\bm{Z}$ with feature dimension $D$:
\begin{align}
   \label{eq:encode}
   \bm{Z} = \mathrm{gLN}(\mathrm{Conv2D}(\bm{X})) \in \R^{D \times T \times F},
\end{align}
where gLN is global layer normalization~\cite{convtasnet}.
For frequency modeling, we view the feature $\bm{Z}$ as a stack of $T$ arrays of shape $D \times F$, where $D$ is the feature dimension and $F$ the sequence length (i.e., we permute the dimension order of $\bm{Z}$ to $T \times D \times F$).
Frequency modeling is then performed as:
\begin{align}
   \label{eq:sequence_modeling}
   \bm{Z} &\xleftarrow{} \bm{Z} + \mathrm{ConvSwiGLU}(\bm{Z}) / 2, \\
   \bm{Z} &\xleftarrow{} \bm{Z} + \mathrm{MHSA}(\mathrm{Norm}(\bm{Z})), \\
   \bm{Z} &\xleftarrow{} \bm{Z} + \mathrm{ConvSwiGLU}(\bm{Z}) / 2,
\end{align}
where MHSA stands for multi-head self-attention~\cite{transformer}.
MHSA has $H$ heads and each head processes $D/H$-dimensional feature.
We use rotary positional encoding~\cite{rope} for encoding the relative position of each frequency bin.
The $\mathrm{ConvSwiGLU}$ module and the $\mathrm{Norm}$ layer will be described in Sections~\ref{ssec:conv_swiglu} and \ref{ssec:group_norm}, respectively.
Temporal modeling is done in the same way by viewing the feature $\bm{Z}$  as a stack of $F$ arrays of shape $D \times T$, where $T$ is the sequence length (i.e., we permute the dimension order of $\bm{Z}$ to $F \times D \times T$).
After alternating between frequency and temporal modeling $B$ times, the final feature $\bm{Z}$ is used to estimate the RI components of $N$ sources:
\begin{align}
   \label{eq:decode}
   \hat{\bm{S}} = \mathrm{DeConv2D}(\bm{Z}) \in \R^{2 \times N \times T \times F}.
\end{align}
Finally, we obtain the time-domain signals by inverse STFT.

\subsection{ConvSwiGLU module}
\label{ssec:conv_swiglu}

RNN-based models have shown strong performance in TF-domain dual-path models~\cite{tfgridnet}, which
we attribute to the inherent ability of their hidden state updates to better capture local information. 
In contrast, it is expected to be hard for the MHSA to always have locally smooth attention weights.
This consideration motivates us to design a model that has strong local-modeling capability.

To this end, we introduce a new FFN, named ConvSwiGLU.
ConvSwiGLU boosts the local-modeling capability by utilizing 1d-convolution and 1d-deconvolution layers instead of linear layers.
We also exploit the SwiGLU activation, which has shown better performance than the Swish activation in the NLP field~\cite{swiglu}.
Formally, its processing is written as:
\begin{align}
   \label{eq:conv_swiglu}
   \bm{Z} &\xleftarrow{} \mathrm{Norm}(\bm{Z}), \\
   \bm{Z} &\xleftarrow{} \mathrm{Swish}(\mathrm{Conv1D}(\bm{Z})) \otimes \mathrm{Conv1D}(\bm{Z}), \label{eq:swiglu}\\
   \bm{Z} &\xleftarrow{} \mathrm{Deconv1D}(\bm{Z}).
\end{align}
As described in Section~\ref{ssec:overview}, we place this ConvSwiGLU both before and after the self-attention, which enables stronger local modeling.

\subsection{Group normalization for TF-domain dual-path models}
\label{ssec:group_norm}

Typically, TF-domain dual-path models utilize layer normalization or root mean square normalization (RMSNorm)~\cite{rmsnorm} as $\mathrm{Norm}$ layer, to normalize the $D$-dimensional vector of each TF bin.
However, since the goal is the separation of multiple speakers, we believe that it may be beneficial to encourage each $D$-dimensional vector $\bm{Z}_{t,f}$ to be split into groups corresponding to disentangled concepts such as speaker IDs.

In RMSGroupNorm, we view each $D$-dimensional vector $\bm{Z}_{t,f}$ as a stack of $G$ vectors of dimension $D/G$, where $G$ is the group size, and we normalize each $D/G$-dimensional vector separately.
This encourages the model to disentangle each $D$-dimensional vector into some groups, which could be helpful for speech separation.
Note that we normalize each TF bin, unlike the group normalization in image processing~\cite{groupnorm}.
As in RMSNorm, RMSGroupNorm features an affine transform with two $D$-dimensional learnable parameters.
Note that $G=1$ corresponds to the original RMSNorm.
In experiments, we demonstrate that RMSGroupNorm gives slightly but consistently better performance than RMSNorm in our model.

\begin{table}[t]
\begin{center}
\vspace{-2mm}
\caption{
    Summary of hyper-parameter notations and default model configurations for three model sizes: Small (S), Medium (M), and Large (L).
}
\vspace{-3mm}
\label{table:parameters}
\resizebox{\linewidth}{!}{
\begin{tabular}{clrrr}
\toprule

{Symbol} & {Description} & {S} & {M} & {L} \\

\midrule

    $D$ & Embedding dimension of each TF bin & 96 & 128 & 128 \\
    $B$ & Number of Locoformer blocks & 4 & 6 & 9 \\

\midrule

    $C$ & Hidden dimension in Conv-SwiGLU & 256 & 384 & 384 \\
    $K$ & Kernel size in Conv1D and Deconv1D & 4 & 4 & 4 \\
    $S$ & Stride in Conv1D and Deconv1D & 1 & 1 & 1 \\
    $H$ & Number of heads in self-attention & 4 & 4 & 4 \\
    $G$ & Number of groups in RMSGroupNorm & 4 & 4 & 4 \\

\midrule

    -   & Number of parameters [M] & 5.0 & 15.0 & 22.5 \\
     
\bottomrule

\end{tabular}
}
\end{center}
\vspace{-4mm}
\end{table}

\section{Experiments}
\label{sec:experiments}

\subsection{Dataset and experimental setup}
\label{ssec:setup}

To evaluate the model, we used three speech separation corpora, WSJ0-2mix~\cite{dc}, Libri2mix~\cite{librimix}, and WHAMR!~\cite{whamr}, and a speech enhancement corpus, the Interspeech DNS-challenge 2020 dataset (denoted as DNS)~\cite{dns2020}.
In the speech separation datasets, we always used the fully overlapped \texttt{min} version with a sampling rate of 8~kHz, while the sampling rate of the DNS dataset was 16~kHz.

\noindent\textbf{WSJ0-2mix} contains two-speaker mixtures of utterances from the WSJ0 corpus.
The total lengths of the training, validation, and test sets are 30~h, 10~h, and 5~h, respectively.

\noindent\textbf{Libri2Mix} contains two-speaker mixtures of utterances from Librispeech~\cite{librispeech}.
The total lengths of the training, validation, and test sets are 212~h, 11~h, and 11~h, respectively.

\noindent\textbf{WHAMR!} is a noisy reverberant version of WSJ0-2mix. %
The model is trained to jointly perform dereverberation, denoising, and separation.

\noindent\textbf{DNS} has 2700~h of training data and 300~h of validation data, which are simulated using the official script~\cite{dns2020}. %
The non-blind anechoic test set is used for testing.

All experiments are done using the ESPnet-SE pipeline~\cite{espnet_se}.
A summary of hyper-parameters and the default model configurations are shown in Table~\ref{table:parameters}.
We mainly investigate the Medium model.
On some datasets, we also evaluate the Small and Large models to show fairer comparisons with previous models.
Only on WHAMR!, which has strong reverberation, we set $K=8$ and halve $C$, mimicking TF-GridNet. %
We set the STFT window and hop sizes to 16~\si{\milli\second} and 8~\si{\milli\second} for all datasets, except for WHAMR! where the window size is set to 32~\si{\milli\second}.
We use the AdamW optimizer~\cite{adamw} with a weight decay of 1e-2.
We first linearly increase the learning rate from 0 to 1e-3 over the first 4000 training steps.
The learning rate is then decayed by 0.5 if the validation loss does not improve for 3 epochs.
The training is conducted for up to 150 epochs\footnote{On the DNS dataset, we define 21,000 training steps as one epoch because the training data is very large.} with early stopping if the validation loss does not improve for 10 epochs.
When using dynamic mixing (DM), training is done up to 200 epochs, and the learning-rate decay and early stopping are applied after 75 epochs in the Small and Medium models, and 65 epochs in the Large model.
The batch size is 4 and input samples are 4-second long.
The $L_2$ norm of the gradient is clipped to 5.
Each input mixture is normalized by dividing it by its standard deviation.

We use permutation-invariant~\cite{dc, pit} scale-invariant signal-to-noise ratio (SI-SNR)~\cite{sisdr} as the loss function for speech separation.
For speech enhancement, we used a time-domain $L_1$ loss plus a TF-domain multi-resolution $L_1$ loss~\cite{si_mrl1}.
We used four FFT window sizes $\{ 256, 512, 768, 1024 \}$ with 50\% overlap.

In evaluation, we use as weights the averages over the five model checkpoints that had the best losses on the validation set.
We use the following evaluation metrics: SI-SNR improvement (SI-SNRi), signal-to-distortion ratio improvement (SDRi)~\cite{SDR-Vincent2006}, short-time objective intelligibility (STOI)~\cite{stoi}, and wide-band perceptual evaluation of speech quality (PESQ-WB)~\cite{pesq}.

\subsection{Anechoic speech separation and enhancement}
\label{ssec:results_anechoic}

\begin{table}[t]
\sisetup{
detect-weight, %
mode=text, %
tight-spacing=true,
round-mode=places,
round-precision=1,
table-format=2.1,
table-number-alignment=center
}
\begin{center}
\vspace{-2mm}
\caption{
    Comparison with previous models on WSJ0-2mix. Methods with $^*$ use speed perturbation when doing dynamic mixing. ``-'' denotes unavailable result in original work. Results in [dB].
}
\vspace{-3mm}
\label{table:wsj0-2mix}
\resizebox{\linewidth}{!}{
\begin{tabular}{lcS[table-format=3.1]*{4}{S}}
\toprule
& & 
& \multicolumn{2}{c}{w/o DM} & \multicolumn{2}{c}{w/ DM} \\

\cmidrule(lr){4-5}\cmidrule(lr){6-7}

{System} & {\hspace{-.1cm}Domain\hspace{-.1cm}} & 
 {\hspace{-.1cm}\#params[M]\hspace{-.1cm}} 
& {SI-SNRi} & {SDRi} & {SI-SNRi} & {SDRi}  \\

\midrule
    DPRNN~\cite{dprnn} & T &2.6 &18.8 &19.0 &{-} &{-}  \\ 
    
    DPTNet~\cite{dptnet} & T &2.7 &20.2 &20.6 &{-} &{-} \\ 

    Wavesplit~\cite{wavesplit} & T &29.0 &21.0 &21.2 &22.2 &22.3  \\ 

    SepFormer$^*$~\cite{sepformer} & T &25.7 &20.4 &20.5 &22.3 &22.4  \\ 

    TFPSNet~\cite{tfpsnet} & TF &2.7 &21.1 &21.3 &{-} &{-}  \\ 

    QDPN$^*$~\cite{qdpn} & T &200.0 &22.1 &{-} &23.6 &{-}  \\ 

    TF-GridNet~\cite{tfgridnet} & TF &14.4 &23.5 &23.6 &{-} &{-}  \\ 

    MossFormer2$^*$~\cite{mossformer2}\hspace{-.6cm} & T &55.7 &{-} &{-} &24.1 &{-}  \\ 

    SepTDA$_2$~\cite{septda} & T &21.2 &24.0 &23.9 &{-} &{-}  \\ 

\midrule

    TF-Locoformer (S)\hspace{-.2cm} & TF &5.0  &22.0 &22.1 &22.8 &23.0  \\

    TF-Locoformer (M)\hspace{-.2cm} & TF &15.0 &23.6 &23.8 &24.6 &24.7  \\

    TF-Locoformer (L)\hspace{-.2cm} & TF &22.5 &\bfseries 24.2 &\bfseries 24.3 &\bfseries 25.1 &\bfseries 25.2  \\

\bottomrule

\end{tabular}
}
\end{center}
\vspace{-6mm}
\end{table}

\begin{figure}[t]
\centering
\vspace{0mm}
\centerline{\includegraphics[width=\linewidth, trim={.05cm 0 .25cm .4cm},clip]{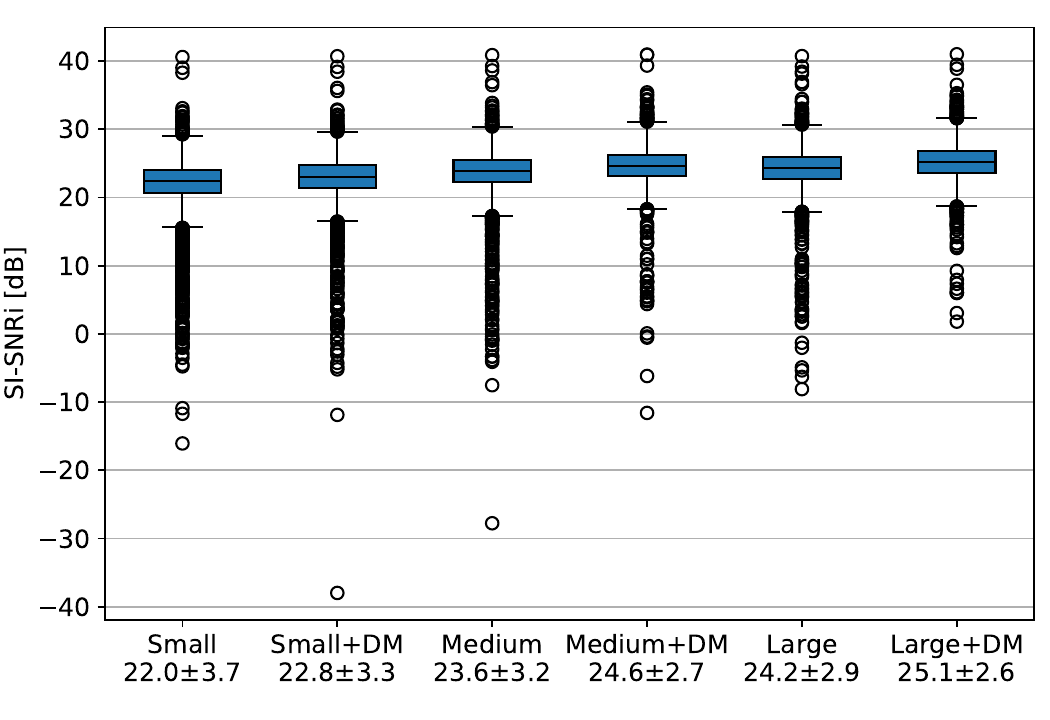}}
\vspace{-4mm}
\caption{
    Box-plots of SI-SNRi [dB] for models with different sizes on WSJ0-2mix test set.
    The numbers below the model size indicate average and standard deviations of SI-SNRi.
}
\label{fig:boxplot}
\vspace{-2mm}
\end{figure}

\begin{table}[t]
\sisetup{
detect-weight, %
mode=text, %
tight-spacing=true,
round-mode=places,
round-precision=1,
table-format=2.1,
table-number-alignment=center
}
\vspace{-2mm}
\begin{center}
\caption{
    Comparison with previous models on Libri2Mix. DM and speed perturbation were not used. Results in [dB].
}
\vspace{0mm}
\label{table:libri2mix}
\resizebox{0.98\linewidth}{!}
{
\begin{tabular}{lcS[table-format=3.1]*{2}{S}}
\toprule

{System} & {Domain} & {\hspace{-.1cm}\#params[M]\hspace{-.1cm}} & {SI-SNRi} & {SDRi}  \\

\midrule
    Conv-TasNet~\cite{convtasnet} & T &5.1 &14.7 &{-}  \\ 
    
    Wavesplit~\cite{wavesplit} & T &29.0 &19.5 &20.0   \\ 

    SepFormer~\cite{sepformer} & T &25.7 &19.2 &19.4  \\ 

    MossFormer2~\cite{mossformer2}\hspace{-.3cm} & T &55.7 &21.7 &{-}  \\ 

\midrule
    TF-Locoformer (M) & TF &15.0 &\bfseries 22.1 & \bfseries22.2  \\

\bottomrule

\end{tabular}
}
\end{center}
\vspace{-5mm}
\end{table}

\begin{table}[t]
\sisetup{
detect-weight, %
mode=text, %
tight-spacing=true,
round-mode=places,
round-precision=1,
table-format=2.1,
table-number-alignment=center
}
\begin{center}
\vspace{0mm}
\caption{
    Comparison with previous non-causal models on the DNS2020 non-blind test dataset. SI-SNR results in [dB].
}
\vspace{-3mm}
\label{table:dns20}
\resizebox{\linewidth}{!}{
\begin{tabular}{lS[table-format=2.1]*{2}{S}S[table-format=1.2,round-precision=2]}
\toprule

{System} & {\hspace{-.1cm}\#params[M]\hspace{-.1cm}} & {SI-SNR} & {STOI} & {PESQ-WB}  \\

\midrule
    Noisy &{-} &9.1 &91.5 & 1.58  \\ 
    
    MFNet~\cite{mfnet} &6.1 &20.3 &98.0 &3.43  \\ 

    USES~\cite{uses} &3.1 &21.2 &98.1 &3.46  \\ 

\midrule
    TF-Locoformer (M) &15.0 &\bfseries 23.3 & \bfseries 98.8 & \bfseries 3.72  \\

\bottomrule

\end{tabular}
}
\end{center}
\vspace{-2mm}
\end{table}

\begin{table}[t]
\sisetup{
detect-weight, %
mode=text, %
tight-spacing=true,
round-mode=places,
round-precision=1,
table-format=2.1,
table-number-alignment=center
}
\begin{center}
\vspace{-5mm}
\caption{
    Comparison with previous models on WHAMR!. 
    DM$^*$ indicates DM with speed perturbation. Results in [dB].
}
\vspace{-3mm}
\label{table:whamr}
\resizebox{\linewidth}{!}{
\begin{tabular}{lcS[table-format=3.1]SS}
\toprule

{System} & {Domain } & {\hspace{-.1cm}\#params[M]\hspace{-.1cm}} & {SI-SNRi} & {SDRi}  \\

\midrule
    
    Wavesplit+DM~\cite{wavesplit} & T &29.0 &13.2 &12.2   \\ 

    SepFormer+DM$^*$~\cite{sepformer} & T &25.7 &14.0 &13.0  \\ 

    QDPN+DM$^*$~\cite{qdpn} & T &200.0 &14.4 &{-}  \\ 

    MossFormer2+DM$^*$~\cite{mossformer2}\hspace{-.3cm} & T &55.7 &17.0 &{-}  \\ 

    TF-GridNet~\cite{tfgridnet} & TF &5.5 &17.1 &15.6  \\ 

\midrule

    TF-Locoformer (S) & TF &5.0 &17.4 &15.9  \\

    TF-Locoformer (M) & TF &15.0 &\bfseries 18.5 &\bfseries 16.9  \\

\bottomrule

\end{tabular}
}
\end{center}
\vspace{-5mm}
\end{table}

Table~\ref{table:wsj0-2mix} compares the performance of the proposed model with the results reported in the literature on the WSJ0-2mix dataset.
First, we focus on the TF-domain models, TFPSNet and TF-GridNet, both of which have RNNs.
We compare them with the Medium model because all of them have $B=6$ dual-path blocks.
The result demonstrates that TF-Locoformer achieves comparable or better performance than RNN-based TF-domain models, implying that RNN-free models can work well in the TF-domain by introducing strong local modeling.
Next, we compare TF-Locoformer with the other models.
Since the current SoTA model on WSJ0-2mix, SepTDA, has nine multi-path blocks, the Large model is suitable for this comparison.
TF-Locoformer again achieves comparable or better performance than previous SoTA models.
We also trained the models using dynamic mixing and observed noticeable improvements.

Although it is often argued that the performance on WSJ0-2mix is saturated, we find that there is still room for improvement on samples which give low SI-SNRi.
Figure~\ref{fig:boxplot} shows the boxplots of SI-SNRi given by the proposed models on the WSJ0-2mix test set.
It can be observed that the larger models reduce the number of failures and work more robustly.
Dynamic mixing results in a similar effect by augmenting the data.

On the larger-scale speech separation and enhancement datasets, TF-Locoformer works well too.
Table~\ref{table:libri2mix} and Table~\ref{table:dns20} show the performance of TF-Locoformer and competing models on the Libri2Mix and DNS datasets, respectively.
The proposed TF-Locoformer gives the best performance on both datasets, which demonstrates its potential scalability.
In addition, TF-Locoformer is also effective for denoising.

\setcounter{topnumber}{3}

\subsection{Noisy reverberant speech separation}
\label{ssec:results_reverb}

Table~\ref{table:whamr} compares the performance of the proposed model with previous models on the WHAMR! dataset.
We evaluated the Small model to fairly compare with the current SoTA model, TF-GridNet, because it was configured to have $B=4$ dual-path blocks on WHAMR! evaluation~\cite{tfgridnet}.
The result shows that the proposed model again outperforms the SoTA models.
The Medium model scores even higher, implying that the proposed model could still achieve even better performance by using larger configurations.

The comparison not only shows that TF-Locoformer is effective, but also demonstrates that the current best TF-domain models perform better than the best available time-domain approaches in reverberant conditions.
While some models such as QDPN~\cite{qdpn} or MossFormer2~\cite{mossformer2} gave comparable or better performance than the Small TF-Locoformer on WSJ0-2mix, the latter achieves better performance on WHAMR!, even without dynamic mixing.
This observation is in line with~\cite{cord2022monaural}, which argues that time-domain models struggle with reverberation due to their short kernel size in the encoder and decoder (e.g, 2~\si{\milli\second}).
The results clearly show the superiority of TF-domain models in reverberant conditions, which are more representative of real-world applications.

\subsection{Ablation study}
\label{ssec:ablation_study}

\begin{figure}[t]
\centering
\centerline{\includegraphics[width=0.825\linewidth]{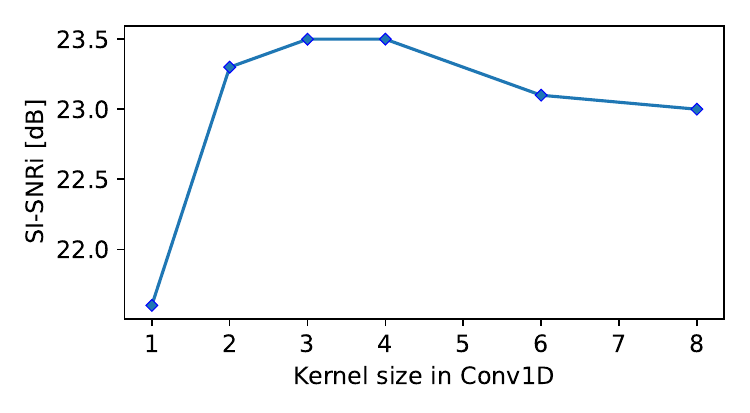}}
\vspace{-4mm}
\caption{
    Average SI-SNRi with different kernel sizes on WSJ0-2mix test set.
    Medium model is shown.
}
\label{fig:kernel_size}
\vspace{-4mm}
\end{figure}
\begin{table}[t]
\begin{center}
\caption{
    Ablation study on WSJ0-2mix. Results in [dB].
}
\vspace{-3mm}
\label{table:ablation}
\resizebox{\linewidth}{!}{
\begin{tabular}{llrr}
\toprule

& {System} & {SI-SDRi} & {SDRi} \\

\midrule
    {\texttt{A0}} & TF-Locoformer (M) &\bfseries 23.6 & \bfseries 23.8  \\

    {\texttt{A1}} & ~~ Macaron-style $\xrightarrow{}$ Single ConvSwiGLU & 22.8 & 22.9  \\

    {\texttt{A2}} & ~~~~ SwiGLU $\xrightarrow{}$ Swish activation & 22.2 & 22.4 \\
    
\bottomrule

\end{tabular}
}
\end{center}
\vspace{0mm}
\end{table}

\begin{table}[t]
\sisetup{
detect-weight, %
mode=text, %
tight-spacing=true,
round-mode=places,
round-precision=1,
table-format=2.1,
table-number-alignment=center
}
\begin{center}
\vspace{-7mm}
\caption{
    Comparison of normalization layers on WSJ0-2mix. Results in [dB].
}
\vspace{0mm}
\label{table:groupnorm}
\resizebox{0.9\linewidth}{!}{
\begin{tabular}{l*{4}{S}}
\toprule
& \multicolumn{2}{c}{RMSNorm} & \multicolumn{2}{c}{RMSGroupNorm} \\

\cmidrule(lr){2-3}\cmidrule(lr){4-5}

{System} & {SI-SNRi} & {SDRi} & {SI-SNRi} & {SDRi}  \\

\midrule

    TF-Locoformer (S) &21.7 &21.9 &\bfseries 22.0 &\bfseries 22.1  \\

    TF-Locoformer (M) &23.5 &23.6 &\bfseries 23.6 &\bfseries 23.8  \\

    TF-Locoformer (L) &24.0 &24.1 &\bfseries 24.2 &\bfseries 24.3  \\

\bottomrule

\end{tabular}
}
\end{center}
\vspace{-8mm}
\end{table}

We conducted an ablation study to examine the effectiveness of each module in our model: convolution layers, ConvSwiGLU FFN, Macaron-style architecture, and RMSGroupNorm.
Figure~\ref{fig:kernel_size} shows the average SI-SNRi of the Medium models with different kernel sizes on the WSJ0-2mix test set.
Since the default model configuration was $K\thineq 4$ and $C\thineq 384$, we set $C \thineq 1536 / K$ to make all the models roughly the same size.
We used here the RMSNorm as the normalization layer to avoid potential influence of RMSGroupNorm's grouping on the results.
$K\thineq 1$ is equivalent to a normal linear layer, in which case each block is almost the same as the pure Transformer block.
The results clearly demonstrate the importance of the convolution layer: the models with $K \geq 2$ perform much better than that with $K=1$.
At the same time, models with longer kernels are more computationally efficient because the input to the Swish activation and the following gating have smaller hidden dimension $C$.
However, we observe a trade-off between efficiency and performance: models with longer kernels (e.g., $K=8$) do not give the best performance.
Instead, we find that $K=3$ and $K=4$ lead to the best results.

In Table~\ref{table:ablation}, we evaluate the contribution of our other modifications to the original Transformer block while keeping the model size constant.
\texttt{A1} removes the first ConvSwiGLU module while increasing the hidden dimension in the second ConvSwiGLU to $2C$. %
\texttt{A2} further swaps the SwiGLU activation for Swish (i.e., removes the right Conv1D branch in Fig.~\ref{fig:overview}) while increasing the hidden dimension to $3C$.
The result demonstrates that both designs help improving the separation performance. 

Finally, we compare the performance of the models with RMSNorm and RMSGroupNorm in Table~\ref{table:groupnorm}.
Although the improvement is slight, the proposed RMSGroupNorm consistently leads to better performance.
This result demonstrates that, as discussed in Section~\ref{ssec:group_norm}, encouraging the model to make groups in each TF bin can be effective for TF-domain dual-path separation models.

\section{Conclusion}
\label{sec:conclusion}

We presented TF-Locoformer, a speech separation model that effectively integrates global and local modeling for TF-domain dual-path modeling.
While previous SoTA TF-domain dual-path models are based on RNNs, we developed a model based on the Transformer block, considering its advantages such as parallelizable architecture, potential scalability, and versatility (e.g., prompting to specify the task).
Inspired by the success of RNNs, which perform both local and global modeling, we designed a model where self-attention does global modeling and convolution handles local modeling.
We also proposed an effective normalization layer for TF-domain models.
The experimental comparison demonstrated that the proposed model gives comparable or better performance than previous SoTA models on four commonly-used benchmarks.
Through the ablation study, we have shown the importance of local modeling.
Finally, the experiments implied that TF-domain models deal with reverberation much better than time-domain models.
In the future, we will investigate the scalability of TF-Locoformer, as well as its effectiveness on music and general sound separation. %

\clearpage
\balance

\bibliographystyle{IEEEtran}
\bibliography{main}

\begin{thebibliography}{10}
\providecommand{\url}[1]{#1}
\csname url@samestyle\endcsname
\providecommand{\newblock}{\relax}
\providecommand{\bibinfo}[2]{#2}
\providecommand{\BIBentrySTDinterwordspacing}{\spaceskip=0pt\relax}
\providecommand{\BIBentryALTinterwordstretchfactor}{4}
\providecommand{\BIBentryALTinterwordspacing}{\spaceskip=\fontdimen2\font plus
\BIBentryALTinterwordstretchfactor\fontdimen3\font minus
  \fontdimen4\font\relax}
\providecommand{\BIBforeignlanguage}[2]{{%
\expandafter\ifx\csname l@#1\endcsname\relax
\typeout{** WARNING: IEEEtran.bst: No hyphenation pattern has been}%
\typeout{** loaded for the language `#1'. Using the pattern for}%
\typeout{** the default language instead.}%
\else
\language=\csname l@#1\endcsname
\fi
#2}}
\providecommand{\BIBdecl}{\relax}
\BIBdecl

\bibitem{dc}
J.~R. Hershey, Z.~Chen, J.~Le~Roux, and S.~Watanabe, ``Deep clustering:
  Discriminative embeddings for segmentation and separation,'' in \emph{Proc.
  ICASSP}, 2016.

\bibitem{convtasnet}
Y.~Luo and N.~Mesgarani, ``{Conv-TasNet}: Surpassing ideal time-frequency
  magnitude masking for speech separation,'' \emph{IEEE/ACM Trans. Audio,
  Speech, Lang. Process.}, vol.~27, no.~8, pp. 1256--1266, 2019.

\bibitem{dprnn}
Y.~Luo, Z.~Chen, and T.~Yoshioka, ``{Dual-Path RNN}: Efficient long sequence
  modeling for time-domain single-channel speech separation,'' in \emph{Proc.
  ICASSP}, 2020.

\bibitem{dptnet}
J.~Chen, Q.~Mao, and D.~Liu, ``{Dual-Path Transformer Network}: Direct
  context-aware modeling for end-to-end monaural speech separation,'' in
  \emph{Proc. Interspeech}, 2020.

\bibitem{sepformer}
C.~Subakan, M.~Ravanelli, S.~Cornell, M.~Bronzi, and J.~Zhong, ``Attention is
  all you need in speech separation,'' in \emph{Proc. ICASSP}, 2021.

\bibitem{tfpsnet}
L.~Yang, W.~Liu, and W.~Wang, ``{TFPSNet}: Time-frequency domain path scanning
  network for speech separation,'' in \emph{Proc. ICASSP}, 2022.

\bibitem{tfgridnet}
Z.-Q. Wang, S.~Cornell, S.~Choi, Y.~Lee, B.-Y. Kim \emph{et~al.},
  ``{TF-GridNet}: Integrating full-and sub-band modeling for speech
  separation,'' \emph{IEEE/ACM Trans. Audio, Speech, Lang. Process.}, vol.~31,
  pp. 3221--3236, 2023.

\bibitem{cord2022monaural}
T.~Cord-Landwehr, C.~Boeddeker, T.~Von~Neumann, C.~Zoril{\u{a}}, R.~Doddipatla
  \emph{et~al.}, ``Monaural source separation: From anechoic to reverberant
  environments,'' in \emph{Proc. IWAENC}, 2022.

\bibitem{transformer}
A.~Vaswani, N.~Shazeer, N.~Parmar, J.~Uszkoreit, L.~Jones \emph{et~al.},
  ``Attention is all you need,'' \emph{Proc. NeurIPS}, 2017.

\bibitem{qdpn}
J.~Rixen and M.~Renz, ``{QDPN} - quasi-dual-path network for single-channel
  speech separation,'' in \emph{Proc. Interspeech}, 2022.

\bibitem{mossformer2}
S.~Zhao, Y.~Ma, C.~Ni, C.~Zhang, H.~Wang \emph{et~al.}, ``{MossFormer2}:
  Combining transformer and {RNN}-free recurrent network for enhanced
  time-domain monaural speech separation,'' in \emph{Proc. ICASSP}, 2024.

\bibitem{karita2019comparative}
S.~Karita, N.~Chen, T.~Hayashi, T.~Hori, H.~Inaguma \emph{et~al.}, ``A
  comparative study on {T}ransformer vs {RNN} in speech applications,'' in
  \emph{Proc. ASRU}, 2019.

\bibitem{vit}
A.~Dosovitskiy, L.~Beyer, A.~Kolesnikov, D.~Weissenborn, X.~Zhai \emph{et~al.},
  ``An image is worth 16x16 words: Transformers for image recognition at
  scale,'' in \emph{Proc. ICLR}, 2021.

\bibitem{whisper}
A.~Radford, J.~W. Kim, T.~Xu, G.~Brockman, C.~McLeavey \emph{et~al.}, ``Robust
  speech recognition via large-scale weak supervision,'' in \emph{Proc. ICML},
  2023.

\bibitem{uses}
W.~Zhang, K.~Saijo, Z.-Q. Wang, S.~Watanabe, and Y.~Qian, ``Toward universal
  speech enhancement for diverse input conditions,'' in \emph{Proc. ASRU},
  2023.

\bibitem{scaling_law}
J.~Kaplan, S.~McCandlish, T.~Henighan, T.~B. Brown, B.~Chess \emph{et~al.},
  ``Scaling laws for neural language models,'' \emph{arXiv preprint
  arXiv:2001.08361}, 2020.

\bibitem{scaling_law_acoustic_model}
J.~Droppo and O.~Elibol, ``Scaling laws for acoustic models,'' in \emph{Proc.
  Interspeech}, 2021.

\bibitem{scaling_law_vit}
X.~Zhai, A.~Kolesnikov, N.~Houlsby, and L.~Beyer, ``Scaling vision
  transformers,'' in \emph{Proc. CVPR}, 2022.

\bibitem{conformer}
A.~Gulati, J.~Qin, C.-C. Chiu, N.~Parmar, Y.~Zhang \emph{et~al.}, ``Conformer:
  Convolution-augmented {T}ransformer for speech recognition,'' \emph{arXiv
  preprint arXiv:2005.08100}, 2020.

\bibitem{macaron_net}
Y.~Lu, Z.~Li, D.~He, Z.~Sun, B.~Dong \emph{et~al.}, ``Understanding and
  improving transformer from a multi-particle dynamic system point of view.''
  in \emph{ICLR 2020 Workshop on Integration of Deep Neural Models and
  Differential Equations}, 2020.

\bibitem{rope}
J.~Su, Y.~Lu, S.~Pan, B.~Wen, and Y.~Liu, ``Ro{F}ormer: Enhanced transformer
  with rotary position embedding,'' \emph{arXiv preprint arXiv:2104.09864},
  2021.

\bibitem{swiglu}
N.~Shazeer, ``{GLU} variants improve {T}ransformer,'' \emph{arXiv preprint
  arXiv:2002.05202}, 2020.

\bibitem{rmsnorm}
B.~Zhang and R.~Sennrich, ``Root mean square layer normalization,'' \emph{Proc.
  NeurIPS}, 2019.

\bibitem{groupnorm}
Y.~Wu and K.~He, ``Group normalization,'' in \emph{Proc. ECCV}, 2018.

\bibitem{librimix}
J.~Cosentino, M.~Pariente, S.~Cornell, A.~Deleforge, and E.~Vincent,
  ``Libri{M}ix: An open-source dataset for generalizable speech separation,''
  \emph{arXiv preprint arXiv:2005.11262}, 2020.

\bibitem{whamr}
M.~Maciejewski, G.~Wichern, E.~McQuinn, and J.~Le~Roux, ``{WHAMR!}: Noisy and
  reverberant single-channel speech separation,'' in \emph{Proc. ICASSP}, 2020.

\bibitem{dns2020}
C.~K. Reddy, V.~Gopal, R.~Cutler, E.~Beyrami, R.~Cheng \emph{et~al.}, ``The
  {INTERSPEECH} 2020 deep noise suppression challenge: Datasets, subjective
  testing framework, and challenge results,'' in \emph{Proc. Interspeech},
  2020.

\bibitem{librispeech}
V.~Panayotov, G.~Chen, D.~Povey, and S.~Khudanpur, ``Libri{S}peech: an {ASR}
  corpus based on public domain audio books,'' in \emph{Proc. ICASSP}, 2015.

\bibitem{espnet_se}
C.~Li, J.~Shi, W.~Zhang, A.~S. Subramanian, X.~Chang \emph{et~al.},
  ``{ESPnet-SE}: End-to-end speech enhancement and separation toolkit designed
  for asr integration,'' in \emph{Proc. SLT}, 2021.

\bibitem{adamw}
I.~Loshchilov and F.~Hutter, ``Decoupled weight decay regularization,'' in
  \emph{Proc. ICLR}, 2018.

\bibitem{pit}
D.~Yu, M.~Kolbæk, Z.~H. Tan, and J.~Jensen, ``Permutation invariant training
  of deep models for speaker-independent multi-talker speech separation,'' in
  \emph{Proc. ICASSP}, 2017.

\bibitem{sisdr}
J.~Le~Roux, S.~Wisdom, H.~Erdogan, and J.~R. Hershey, ``{SDR} --- half-baked or
  well done?'' in \emph{Proc. ICASSP}, 2019.

\bibitem{si_mrl1}
Y.-J. Lu, S.~Cornell, X.~Chang, W.~Zhang, C.~Li \emph{et~al.}, ``Towards
  low-distortion multi-channel speech enhancement: The {ESPNET-SE} submission
  to the {L3DAS22} challenge,'' in \emph{Proc. ICASSP}, 2022.

\bibitem{SDR-Vincent2006}
E.~Vincent, R.~Gribonval, and C.~F{\'e}votte, ``Performance measurement in
  blind audio source separation,'' \emph{IEEE Trans. Audio, Speech, Lang.
  Process.}, vol.~14, no.~4, pp. 1462--1469, 2006.

\bibitem{stoi}
C.~H. Taal, R.~C. Hendriks, R.~Heusdens, and J.~Jensen, ``An algorithm for
  intelligibility prediction of time–frequency weighted noisy speech,''
  \emph{IEEE Trans. Audio, Speech, Lang. Process.}, vol.~19, no.~7, pp.
  2125--2136, 2011.

\bibitem{pesq}
A.~Rix, J.~Beerends, M.~Hollier, and A.~Hekstra, ``Perceptual evaluation of
  speech quality ({PESQ})-a new method for speech quality assessment of
  telephone networks and codecs,'' in \emph{Proc. ICASSP}, 2001.

\bibitem{wavesplit}
N.~Zeghidour and D.~Grangier, ``Wavesplit: End-to-end speech separation by
  speaker clustering,'' \emph{IEEE/ACM Trans. Audio, Speech, Lang. Process.},
  vol.~29, pp. 2840--2849, 2021.

\bibitem{septda}
Y.~Lee, S.~Choi, B.-Y. Kim, Z.-Q. Wang, and S.~Watanabe, ``Boosting
  unknown-number speaker separation with transformer decoder-based attractor,''
  in \emph{Proc. ICASSP}, 2024.

\bibitem{mfnet}
L.~Liu, H.~Guan, J.~Ma, W.~Dai, G.~Wang \emph{et~al.}, ``A mask free neural
  network for monaural speech enhancement,'' in \emph{Proc. Interspeech}, 2023.

\end{thebibliography}
\end{document}